\documentclass[aps,prl,twocolumn,showpacs,superscriptaddress,twoside,showpacs]{revtex4-1}
\pdfoutput=1
\usepackage{graphicx}
\usepackage{dcolumn}
\usepackage{bm}
\usepackage{booktabs}
\usepackage{amsmath}
\usepackage{color}

\begin{document}
\title{Comment on ``Erratum: Negative-parity high-spin states and a possible magnetic rotation band in $^{135}_{59}$Pr$_{76}$ [Phys. Rev. C 92, 054325 (2015)]"}

 \author{S. Guo}
\affiliation{Key Laboratory of High Precision Nuclear Spectroscopy and Center for Nuclear Matter Science, Institute of Modern Physics, Chinese Academy of Sciences, Lanzhou 730000, People's Republic of China} 
\affiliation{School of Nuclear Science and Technology, University of Chinese Academy of Science, Beijing 100049, People's Republic of China}
 \author{C. M. Petrache}
\affiliation{Centre de Sciences Nucl\'eaires et Sciences de la
  Mati\`ere, CNRS/IN2P3, Universit\'{e} Paris-Saclay, B\^at. 104-108, 91405  Orsay, France}

 \date{}
\begin{abstract}

In [Ritika Garg et al., Phys. Rev. C 100, 069901(E) (2019)] the experimental results on the polarization asysmetry were revised due to a claimed change of the geometry asymmetry. However, the revised results can not be reproduced as claimed in the erratum by simply changing the geometry asymmetry in extracting the polarization asymmetry, without re-extracting the polarization asymmetry from the original experimental data. It is possible that the quoted errors were significantly underestimated.

\end{abstract}



\maketitle

In Ref. \cite{Garg} a series of results on polarization asymmetry ($\Delta$) were reported. Among them the results of two transitions, with energies at 747.5 and 813.3 keV, were in contradiction with those reported in Ref. \cite{Matta}. This contradiction gained wide attention since the reported E2 dominating character on these two transitions provided the critical experimental proof for the first wobbling band out of A$\sim$160 mass region. However, the results in Ref. \cite{Garg} published six month later reported a negative polarization assymetry which induces a dominant M1 character of the two transitions contradicting the wobbling interpretation. Note that the two contradicting results were  obtained using the same reaction ($^{16}O+^{123}Sb$) and the same detectors array (INGA).

\begin{figure}[ht]
	\vskip -. cm
	\hskip -. cm
	\centering\includegraphics[clip=true,width=0.4\textwidth]{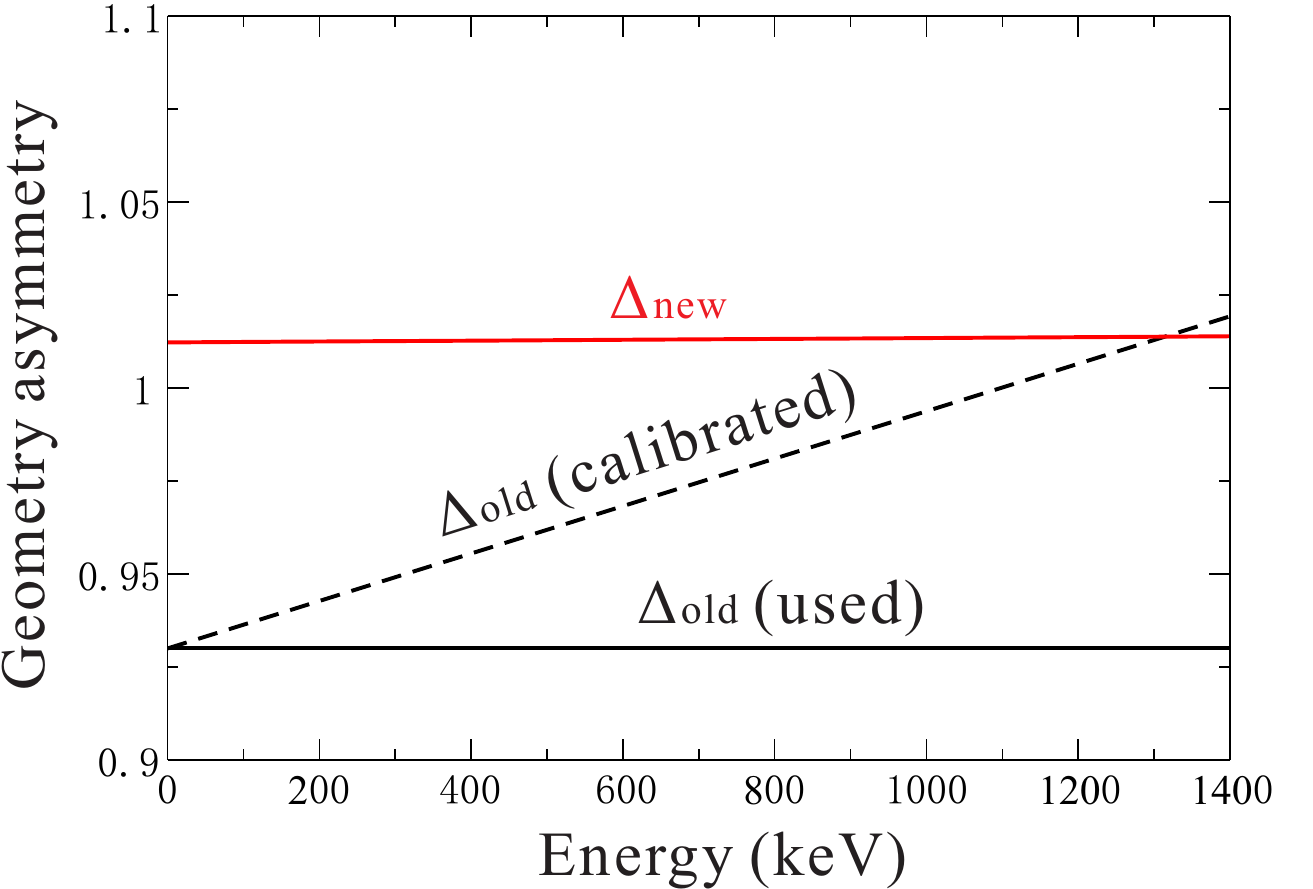}
	\vskip -0. cm
	\caption{(Color online) Geometry asymmetry as functions of transition energy. }
	\label{cal}
\end{figure}

Recently, this contradiction has been resolved by the commented erratum, which revised the results to be in coincidence with Ref. \cite{Matta}. The revision on the results was explained as due to the change of the geometry asymmetry ($a$), which is plotted in Fig. 1. The old $a$ values were calibrated as an increasing line with a considerable slope, though the coefficient $a_1 = 6.38 \times 10^{-5}$ was considered to be insignificance and neglected. On the contrary, the new $a$ values were claimed to be energy dependent, but a small coefficient $a_1 = 1.17083 \times 10^{-6}$ made them quite flat. Therefore both the old and the new actually used $a$ values are close to constants, with a nearly fixed difference ($\sim 0.08$) between them. Comparing to the error bar of the old $a$ values (0.06), the difference is quite large. However, the origin of the difference was not explained.

In the commented erratum, the new results were claimed to be determined by using the new geometry asymmetry, without mentioning any change in the original data. There is a simple relation (see Eq. \ref{e1}, which is quoted from Ref. \cite{Garg}, Eq. 2) among the polarization aymmetry ($\Delta$), geometry asymmetry ($a$) and the ratio (R) between the number of perpendicular ($N_\perp$) and that of parallel ($N_\parallel$) scattering. 

\begin{equation}
\Delta = \frac{a N_\perp-N_\parallel}{a N_\perp+N_\parallel}\label{e1}
\end{equation}

According to Eq. \ref{e1}, R can be deduced when the polarization aymmetry and geometry asymmetry are known:

\begin{equation}
R = \frac{N_\perp}{N_\parallel} = \frac{1+\Delta}{1-\Delta}/a\label{e2}
\end{equation}

Using Eq. \ref{e2}, the old and new R values have been deduced and listed in Table \ref{tab01}, which are expected to keep unchanged when only the geometry asymmetry is changed. The errors are not deduced since the error on the new geometry asymmetry was not reported. Surprisingly, significant change has been found for all transitions. For 747.5- and 813.3-keV transitions, the R values were modified from 0.973 and 0.717 to 1.048 and 1.069, respectively. To make it clear how the new $\Delta$ results would be without inducing any change on R, we use Eq. \ref{e3} which is deduced from Eqs. \ref{e1} and \ref{e2}. The deduced new R values are listed in Table \ref{tab01} and plotted in Fig. \ref{result}. 

\begin{table}
	\renewcommand{\baselinestretch}{1.0}
	
	\caption{$\gamma$-ray energy, polarization asymmetry ($\Delta$) ratios and deduced ratios between the counts of parallel and perpendicular scattering (R).}
	\label{tab01}
	\begin{tabular}{llllll}
		\hline\hline
		
		{$E_{\gamma }$(keV)} &{$\Delta _{old}$ }& {$\Delta _{new}$ }&{$\Delta _{deduced}$ } & {$R_{old}$} & {$R_{new}$}\\
		\hline
		~\\
		
		325.1&	-0.12(6)&	-0.01(3)&	-0.078&	0.845&	0.968\\
		332.9&	-0.03(8)&	-0.01(3)&	0.013&	1.013&	0.968\\
		372.8&	0.08(4)&	0.11(1)&	0.122&	1.262&	1.232\\
		410.8&	-0.16(10)&	-0.11(3)&	-0.118&	0.779&	0.792\\
		424.0&	-0.15(9)&	-0.05(6)&	-0.108&	0.795&	0.893\\
		429.7&	-0.05(6)&	-0.04(4)&	-0.007&	0.973&	0.911\\
		498.5&	-0.14(7)&	-0.13(8)&	-0.098&	0.811&	0.760\\
		593.7&	-0.13(12)&	-0.02(3)&	-0.088&	0.828&	0.948\\
		660.2&	0.06(3)&	0.09(1)&	0.102&	1.213&	1.182\\
		688.8&	0.07(6)&	0.04(4)&	0.112&	1.237&	1.069\\
		747.5&	-0.05(3)&	0.03(3)&	-0.007&	0.973&	1.048\\
		776.2&	0.15(12)&	0.07(8)&	0.192&	1.455&	1.136\\
		813.3&	-0.20(8)&	0.04(3)&	-0.159&	0.717&	1.069\\
		834.0&	0.24(14)&	0.13(4)&	0.280&	1.754&	1.282\\
		854.0&	0.09(6)&	0.13(1)&	0.132&	1.288&	1.282\\
		870.8&	0.12(10)&	0.10(5)&	0.162&	1.369&	1.206\\
		999.9&	0.12(6)&	0.08(3)&	0.162&	1.369&	1.158\\
		1075.2&	0.22(12)&	0.08(5)&	0.261&	1.682&	1.158\\
		1197.4&	0.12(11)&	0.05(4)&	0.162&	1.369&	1.090\\
		1225.9&	0.11(7)&	0.09(5)&	0.152&	1.341&	1.182\\
		1363.7&	0.09(5)&	0.08(5)&	0.133&	1.288&	1.158\\
		\hline

	\end{tabular}
\end{table}

\begin{equation}
\Delta_{new} = \frac{a_{new}(1+\Delta_{old})-a_{old}(1-\Delta_{old})}{a_{new}(1+\Delta_{old})+a_{old}(1-\Delta_{old})}\label{e3}
\end{equation}

The deduced $\Delta$ values for the 747.5- and 813.3-keV transitions obtained by maintaining the R values unchanged are -0.007 and -0.159, respectively. The small negative values still indicate the M1 dominated characters. In Fig. \ref{result} it can be seen that the $\Delta_{deduced}$ values are larger than the $\Delta_{old}$ values by approximately constant difference ($\sim$ 0.04). In fact, from the above equations, we can deduced the difference assuming the R values are unchanged:

\begin{figure}[ht]
	\vskip -. cm
	\hskip -. cm
	\centering\includegraphics[clip=true,width=0.4\textwidth]{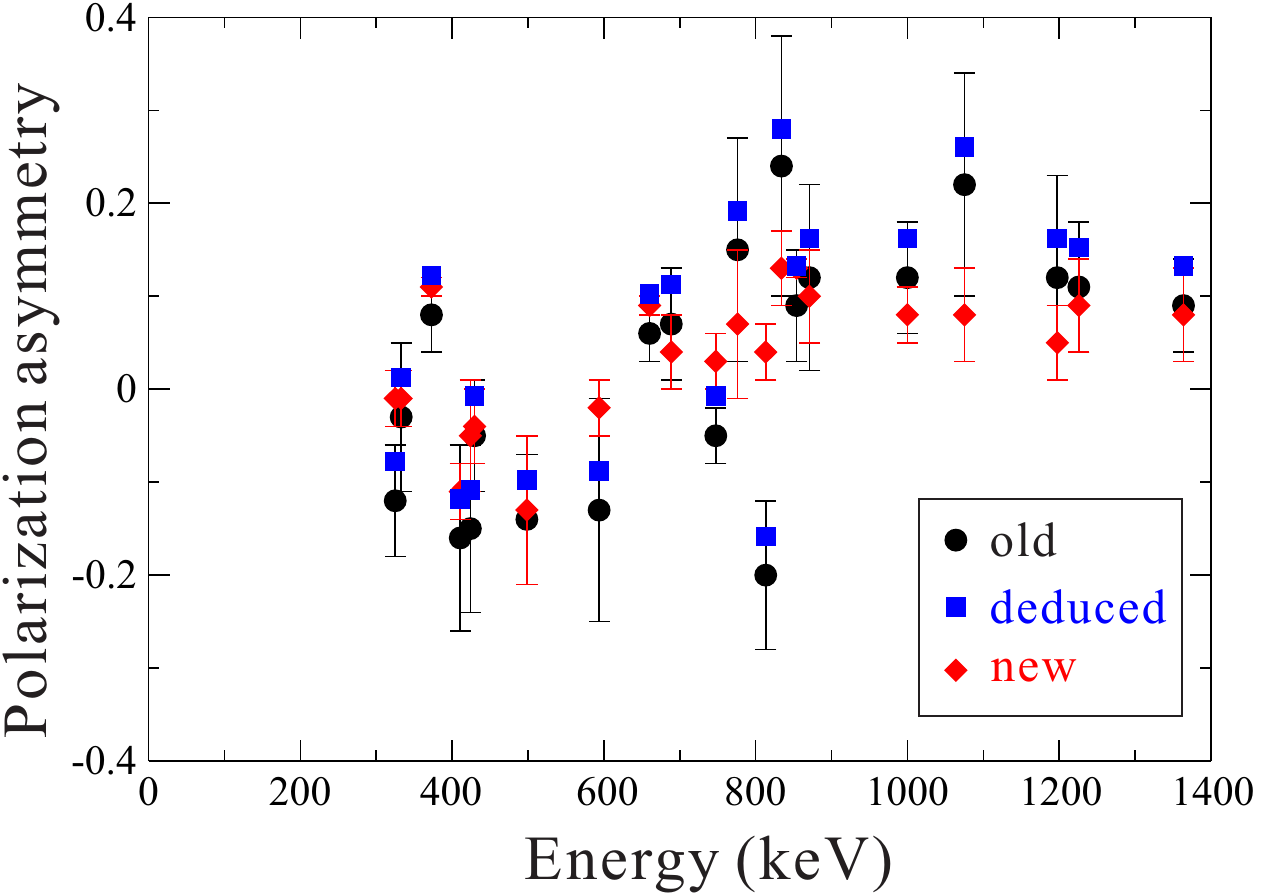}
	\vskip -0. cm
	\caption{(Color online) The old and new polarization asymmetry values, in comparison with the deduced ones assuming only the geometry asymmetry is changed.}
	\label{result}
\end{figure}

\begin{equation}
\Delta_{new} - \Delta_{old} = \frac{(a_{new}-a_{old}) + (a_{new}-a_{old})\Delta_{old}^2}{(a_{new}+a_{old}) + (a_{new}-a_{old}) \Delta_{old}}\label{e4}
\end{equation}

Where $\Delta_{old} \sim 0$, $a_{new} \sim ~1$, and $a_{old} \sim 1$, it approximately equal to:

\begin{equation}
\Delta_{new} - \Delta_{old} \sim \frac{(a_{new}-a_{old})}{2}\label{e5}
\end{equation}

Considering that the new $\Delta$ values published in the erratum and the old $\Delta$ values published in Ref. \cite{Garg} are almost randomly changed (see Fig. \ref{result}), they cannot be achieved by simply changing the geometry asymmetry.

Assuming that no subjective fault is involved, it is possible that the R ratios for the new results were deduced again individually from the original data, not by scaling the former ratios by the new geometry asymmetry. In that case, the errors are significantly underestimated, since the deduced $\Delta$ values using the former original data differ from the new ones too much comparing to the errors. That means, by properly taking into account the statistical and background subtraction induced errors, it is impossible to judge if the M1/E2 characters of the 747.5- and 813.3-keV transitions are M1 or E2 dominated, based on the polarization assymetry deduced from experimental data with statistics at the level of that reported in the work of Ref. \cite{Garg}.
\end{document}